# Field Effect Transistors with Sub-Micrometer Gate Lengths Fabricated from LaAlO$_3$-SrTiO$_3$-Based Heterostructures

C. Woltmann, T. Harada*, H. Boschker, V. Srot, P. A. van Aken, H. Klauk, J. Mannhart

[1]Max Planck Institute for Solid State Research, 70569 Stuttgart, Germany

* present address:
Institute for Materials Research, Tohoku University, Sendai 980-8577, Japan

The possible existence of short-channel effects in oxide field-effect transistors is investigated by exploring field-effect transistors with various gate lengths fabricated from LaAlO$_3$-SrTiO$_3$ heterostructures. The studies reveal the existence of channel-length modulation and drain-induced barrier lowering for gate lengths below 1 µm, with a characteristic behavior comparable to semiconducting devices. With the fabrication of field-effect transistors with gate lengths as small as 60 nm the results demonstrate the possibility to fabricate by electron-beam lithography functional devices based on complex oxides with characteristic lengths of several ten nanometers.

The wide spectrum of functionalities of complex oxides [1,2,3] make these materials highly interesting for use in oxide electronic devices. Devices based on high electronic susceptibility, piezoelectricity, ferroelectricity, resistive switching (RRAM) effects, superconductivity, and sensor functions [4, 5] do exist or are being developed, to give a few examples. For the integration of the functional properties of oxides into electronic systems, field-effect transistors (FETs) or other three-terminal devices based on complex oxides are of value. The viability of such FETs has been explored in the past, using as model system the two-dimensional (2D) electron liquid formed at the $LaAlO_3$-$SrTiO_3$ interface [6] for the drain-source channel. Much work has been focused on this 2D system, as it has been well characterized and because its embedment in high-$K$ materials offers sensitive control by gate fields [7, 8].

One route to fabricating devices based on the 2D $LaAlO_3$-$SrTiO_3$ electron liquid (2DEL) [9] is writing structures by electric fields generated with a charged, conducting tip of a scanning force microscope. This writing technique offers the utmost spatial resolution with line widths as small as 2 nm [10]. While this technique does not allow for top gating, the successful fabrication of field-effect transistors [11, 12], single-electron transistors [13], and a series other devices by using this so-called sketch-based technique has been reported (for an overview see [14]).

Another approach to fabricate $LaAlO_3$-$SrTiO_3$ based devices is to pattern the heterostructures more conventionally by optical lithography, electron-beam lithography, ion irradiation, or using shadow-casting effects as described, *e.g.*, in [15-20]. FETs have been demonstrated with top gates [19, 21-27] as well as with side gates [28]. These transistors operate in a large temperature window which includes room temperature. Also arrays with hundreds of thousands of FETs and monolithic integrated circuits have been fabricated [29]. The large-scale FETs, *i.e.*, FETs with gate lengths $L_G$ of tens or even hundreds of microns which were explored in these studies are characterized by an almost textbook-like behavior, much as standard semiconductor FETs [21].

The characteristics of metal-oxide-semiconductor FETs (MOSFETs) fabricated from canonical semiconductors such as Si or III-V materials much depend on the length of the drain-source channel. With long drain-source channels, MOSFETs are characterized by (*i*) saturation of the drain current $I_D$ at large drain-source voltages $V_D$ for constant gate-source voltages $V_G$; (*ii*) a quadratic dependence of the saturation current $I_{D,sat}$ ($V_G$) for a given $V_D$; and (*iii*) threshold voltages $V_{th}$ being independent of $V_D$ and gate lengths. If the gate lengths are small enough to be comparable to the depletion width [30], these characteristics become dominated by short-channel effects [30, 31] of which the following are particularly important: (a) Channel length modulation caused by a significant fraction of the channel being influenced by the drain potential. Under these conditions, the drain potential reduces the length of the gate's electrostatic barrier, resulting in non-saturating $I_D(V_D)$ characteristics; (b) drain-induced barrier lowering arising when the patterned gate length is reduced below the depletion width and the channel length modulation exceeds a critical value. The drain potential then extends significantly into the channel, additionally reducing the height of the gate's electrostatic barrier. Consequently the channel opens for increasing $V_D$ and the non-saturating $I_D(V_D)$ curves bend to larger current values; (c) velocity saturation occurring due to large longitudinal fields at high drain-source voltages in short channels. Velocity saturation significantly reduces the maximum drain currents in the on-state and linearizes the $I_{D,sat}(V_G)$ characteristic; (d) charge sharing resulting from the source and drain contacts' doping affecting the adjacent regions of the channel. If the extent of the biased regions at source and drain is comparable to the channel length, the channel is turned on more readily. Increasing $V_D$ complements this effect and $V_{th}$ is shifted to lower values for small channel lengths and large $V_D$.

The physics of these short-channel effects is of course controlled by the basic parameters of the FETs, in particular by the carrier densities and dopant concentrations, electric susceptibilities, Fermi velocities, and phonon characteristics of the channels. In addition, the physical properties of the source, drain, and gate contacts play a key role. The complex oxides and the canonical semiconductors much differ in all these fundamental properties [32, 5]. It is therefore an important but open question how the characteristics of FETs built from complex oxides scale at short channel lengths. Do short-channel effects exist, and if so, are they similar to the ones known from semiconducting FETs? In case they exist, at which length scale do they become relevant?

To answer these questions, we have analyzed the behavior of FETs based on LaAlO$_3$-SrTiO$_3$ heterostructures as the gate length is reduced from the range yielding long channel behavior to sub-100 nm lengths. The gated part of the channel is always contacted by the conducting interface and not directly by metal contacts (in contrast to semiconducting FETs, where the gate usually overlaps the metallic contacts). By feeding the source and drain contacts' metallic layers close to both sides of the gate, we strove to keep the series resistance small. Nevertheless, for these devices the gate length is the length relevant for the short-channel effects, rather than the distance between the metal contacts. A cross sectional sketch of these devices is given in Fig. 1. The heterostructures were grown by pulsed laser deposition (PLD) monitored by reflective high-energy electron diffraction (RHEED). First, five unit-cells (uc) of $LaAlO_3$ were epitaxially grown at 800°C and $8\times10^{-5}$ mbar $O_2$ onto a $TiO_2$ terminated (001) surface of a $SrTiO_3$ single-crystal. A $LaAlO_3$ single-crystal was used as target. The 2DEL was patterned as described in [15] using electron-beam lithography. At first, two monolayers $LaAlO_3$ were deposited to protect the interface. Subsequent to the deposition of an amorphous 2DEL negative-pattern, three more $LaAlO_3$ monolayers were added. These $LaAlO_3$ layers generate the 2DEL at the interface and as part of the gate stack acts as gate dielectric. Onto the $LaAlO_3$, 8 monolayers of $BaTiO_3$ were deposited at 660°C and $3\times10^{-3}$ mbar $O_2$. As presented before [29], the $BaTiO_3$ acts as high-$K$ dielectric and is not ferroelectric, likely due to the stress fields surrounding the edge dislocations adapting the lattice mismatch (Fig. 2(b)). The gate stacks were grown *in-situ*, including a top Au contact layer. The gates were patterned *ex-situ* using electron-beam lithography, the Au-Pd bilayer gate electrodes were deposited by PLD and thermal evaporation and patterned by lift-off and Ar-ion milling. To contact the buried electron liquid, 10 nm deep holes were ion-milled through the $BaTiO_3$-$LaAlO_3$ into the $SrTiO_3$, then filled with Ti and capped with Au by using *in-situ* electron-beam evaporation. The heterostructures were characterized by scanning force microscopy and electron microscopy.

To elucidate the local microstructure and composition, cross-sectional cuts through the specimens were analyzed employing scanning transmission electron microscopy (STEM) with atomic-column-resolved high-angle annular dark-field (HAADF) and electron energy-loss (EELS) spectrum imaging. For these studies, the samples were prepared by the *in-situ* lift-out focused ion beam (FIB) technique. These specimens were further thinned to electron transparency by Ar-ion beam milling at low accelerating voltages (200 V). The STEM-HAADF and STEM-EELS measurements were carried out at 200 kV with an advanced analytical TEM/STEM (JEOL ARM200F, JEOL Co. Ltd), equipped with a cold field-emission gun and a DCOR probe Cs-corrector (CEOS Co. Ltd.). EELS elemental maps (2D spectrum images) were obtained in STEM mode with a post-column energy filter with high-speed dual-EELS acquisition capability (Gatan GIF Quantum ERS, Gatan Inc. Pleasanton, USA). The experimental convergence semi-angle was 28 mrad for HAADF and EELS imaging. The corresponding inner and outer collection semi-angles for HAADF were set to 56 and 234 mrad. The inner and outer collection semi-angles for annular dark field (ADF) images acquired simultaneously during EELS spectrum imaging with a Gatan ADF detector were 110 and 270 mrad, and the collection angle for EELS spectrum imaging was 57 mrad. For noise removal from the STEM-HAADF images and -EELS maps a commercially available software package using multivariate statistical analysis (MSA) with weighted principle-component analysis (PCA) for Digital Micrograph (HREM Research Inc.) and a script function written by D.R.G. Mitchell [33] were applied.

The microstructure of the multilayers revealed by these studies matches the design, and the heterostructures are found to be of high quality. The high-angle annular dark-field micrograph showing in cross-sectional view an area underneath the gate of an FET (Fig. 2) reveals that the structure is well epitaxial. Due to the large lattice mismatch between $LaAlO_3$ and $BaTiO_3$ (>5%), misfit dislocations in $BaTiO_3$ associated with an extra atomic plane are frequently observed and accordingly reduce the lattice strain, consistent with previous work [29]. Atomic-column resolved ADF and EELS spectrum images are presented in Fig. 2(c), displaying the atomic arrangement in greater detail, providing further robust evidence of the epitaxial growth of the $BaTiO_3$ (green) and $LaAlO_3$ (red) layers on the $SrTiO_3$ (blue).

Fig. 3(a) provides an optical image of a chip. It contains arrays of several hundred FETs individually connected to bonding pads. All FETs of the chip have the same aspect ratio of the gated area $W_G/L_G = 15$. To compensate for possible alignment offsets during the electron-beam lithography, for each gate length several sets of drain-source distances were implemented. Fig. 3(b) shows a scanning electron microscopy image of an FET with a nominal gate length $L_G = 60$ nm ($W_G = 15 \times 60$ nm = 900 nm) and a drain-source distance of 600 nm. The measurements were all performed at room temperature and in darkness to avoid photo-generated charge carriers. All voltages were applied with respect to the source contact connected to ground. Two source-measure units were each used in two-wire mode to measure the device characteristics.

The output characteristics of devices with $L_G$ = 5 µm, 500 nm, and 60 nm for various gate-source voltages are given in Fig. 4. In this device series, the gates' aspect ratio ($W_G/L_G$ = 15) was kept constant for all devices on the chip. The drain-source distance, however, was not scaled in the same manner due to lithographic constraints. Accordingly, higher resistances act in series to the gated part of the channel for smaller gate lengths. The devices' dimensions are given in table 1, together with their drain-source resistances at vanishing $V_D$ and $V_G$. The transfer characteristics $I_D(V_G)$ of these devices are given in Fig. 5. With the change of the saturation behavior, $I_D(V_D)$ curvature and threshold development, the characteristics show a clear long-channel to short-channel behavior transition of the FETs at $L_G$ ~ 1 µm. The $L_G$ = 5 µm device characteristics (Fig. 4(a)) match the long-channel expectations well. Yet, it is likely that the finite slope in the saturation region is already caused by an onset of channel length modulation. The threshold voltage of this device is $V_G$ = -1.5 V, which matches well the threshold voltage of equivalent FETs that have gate lengths of several ten microns [29]. As shown by Fig. 5 (right plot), the threshold voltage does not depend on $V_D$.

In contrast, the $I_D(V_D)$ characteristics of the $L_G$ = 500 nm FETs (Fig. 4(b)) display a much larger slope in the saturation region, and thereby clearly reveal channel length modulation. Interestingly, the characteristics still feature a soft transition from the linear region ($V_D$ << ($V_G - V_{th}$)) to the saturation region ($V_D > (V_G - V_{th})$). Unfortunately, the possible velocity saturation seems to be hidden by the non-saturating behavior of the curves. The $I_D(V_D)$ characteristics measured with $V_G$ <= -1.4 V display an upturn at large $V_D$, which is a clear sign of drain-induced barrier lowering. $V_{th}$ is shifted to $V_{th}$ < -2 V and is found to vary with $V_D$ (Fig. 5, center panel). This behavior of $V_{th}$ is consistent with the charge-sharing effect as known from conventional semiconductors. No numerical models are available, however, that would predict whether in complex oxides the charge-sharing effect does alter the characteristics in a comparable manner as known from devices built from canonical semiconductors. There is no reason to assume that the short-channel effects in these oxide FETs can simply be extrapolated from the behavior of standard semiconductors. Fig. 4c gives the output characteristics for a device with $L_G$ = 60 nm. Saturation is no longer present and channel length modulation is the dominating effect.

Drain-induced barrier lowering is clearly influencing the $I_D(V_D)$ characteristics for $V_G$ < -2.4 V. The FET's transfer characteristics (Fig. 5, left panel) reveal that $V_{th}$ has shifted to even lower values and is now strongly dependent on $V_D$. Due to channel length modulation and drain-induced barrier lowering, the off-currents of the $L_G$ = 60 nm and $L_G$ = 500 nm devices are about one order of magnitude higher than for the $L_G$ = 5 µm device.

The degradation of the on-state conductivity for decreasing gate lengths in Fig. 5 is attributed to the fact that the fractional length of the ungated channel is larger for devices with small gate lengths. The resulting resistance that is in series to the gated channel is higher and the on-state currents are therefore diminished. Hence, the on/off ratio of the $L_G$ = 60 nm FET (~$10^3$) is much smaller than that of the $L_G$ = 5 µm device (~$10^6$).

Fig. 5 shows the absolute value of the gate current ($|I_G|$) as dashed lines. Gate currents are below $10^{-8}$ A at $V_G$ = -3 V and < $10^{-7}$ A at -9 V (this corresponds to a gate field $E$ ~ 10 MV/cm). We therefore conclude that the influence of $I_G$ on our measurements is negligible.

The device characteristics tend to show drifts on time-scales of weeks. These drifts, as well as the frequently observed hystereses (Fig. 4), are of interest for further studies. The hystereses are significantly smaller for samples that were kept for several hours or days in darkness. The hystereses of the transfer characteristic curves (Fig. 5) increase with decreasing gate length. While this effect needs further investigation, it is possibly caused by hot carrier creation due to the large electric fields generated along short channels. When scattered, some of these carriers are trapped by the gate stack or the substrate, causing hystereses in gate voltage sweeps. In addition, the hystereses may be affected by incompletely quenched ferroelectric behavior of the $BaTiO_3$.

These results provide evidence for the existence of short-channel effects in complex oxide FETs, predominantly comparable to those known from canonical semiconductors. For gate lengths below ~1 µm, channel length modulation, drain-induced barrier lowering and charge sharing were found. The devices are characterized in addition by hysteretic effects usually not observed in semiconducting FETs. Notably, no microscopic model calculations of short-channel effects are available for FETs fabricated from complex oxides. Methods for describing semiconductor physics are applicable only with considerable restrictions, as the electronic properties of the complex oxides differ fundamentally from those of conventional semiconductors. Here we mention in particular non-linear susceptibilities, electronic correlations as well as possibly existing phase transitions of the electron systems. Experiments need to be performed and model calculations to be developed to understand the scaling behavior of complex-oxide field-effect devices at gate lengths below 50 nm, also to explore possible unconventional phenomena controlling the short-channel effects in this class of materials. As in semiconductor devices, the short channel effects will be relevant for possible applications of mesoscopic oxide field effect transistors, and in particular need to be considered in the circuit design of oxide devices.

The authors gratefully acknowledge technical support by U. Waizmann, T. Reindl and J. Weis, and funding by the DFG (Leibniz). The research leading to these results has received funding from the European Union Seventh Framework Program [FP/2007-2013] under grant agreement no 312483 (ESTEEM2).

**Table/Figure Captions:**

**Table 1**

Device dimensions overview. The table gives the devices' gate lengths ($L_G$), gate widths ($W_G$), drain-source distances ($d_{DS}$) and channel resistances ($R_{DS}$) at $V_G = 0$ and vanishing $V_D$. The values of $R_{DS}$ have been extracted from fits to the linear region of the $I_D(V_D)$ curves.

**Figure 1**

Sketch of a cross-section of an FET.

**Figure 2**

High-angle annular dark-field (HAADF) scanning transmission electron microscopy (STEM) cross-sectional images at (a) lower magnification and (b) high magnification of the SrTiO$_3$ substrate, the LaAlO$_3$ and BaTiO$_3$ layers, and the Au capping layer, where a misfit dislocation in BaTiO$_3$ is indicated in (b). (c) Annular dark-field (ADF) STEM image (left) and the respective background-subtracted unfiltered electron energy loss spectroscopy (EELS) elemental maps (Ba green, La red, Sr blue, Ti yellow) of the region underneath the gate electrode. The right image is an RGB overlay of the Ba, La and Sr maps. The scale bar in the ADF image in (c) applies also for the EELS elemental maps.

**Figure 3**

(a) Photograph of a 10 x 10 mm$^2$ sized sample patterned by electron-beam lithography. It comprises several hundred FETs and further test devices. The gate lengths as designed vary from 50 nm to 5 µm.
(b) False-colored scanning electron microscopy image of an FET with 60 nm gate length and a source-drain distance of 600 nm. The 2DEL forms a rectangular region connecting the source and drain electrodes and is surrounded by insulating amorphous LaAlO$_3$. For all FETs on this chip, the aspect ratio of the gated area is the same ($W_G/L_G = 15$).

**Figure 4**

Drain-source transport characteristics of FETs representing three gate length regimes: (a) long channel lengths ($L_G$ = 5 µm), (b) intermediate channel lengths ($L_G$ = 500 nm) with arising short-channel effects and (c) short-channel effect dominated length scales ($L_G$ = 60 nm). The data were taken in dark and at room-temperature.

**Figure 5**

Transfer characteristics of the FETs for the three different gate length regimes shown in Fig. 4. With decreasing gate length the threshold voltage shifts to lower values, and is increasingly depending on the source-drain voltage. Additionally, the on/off ratio and sub-threshold slope decrease. The dashed lines show the absolute value of the gate current. The data were taken in dark and at room-temperature.

**Table 1**

| device number | $L_G$ (µm) | $W_G$ (µm) | $d_{DS}$ (µm) | $R_{DS}$ ($V_G$=0, $V_D$=0) (kΩ) |
|---|---|---|---|---|
| 1 | 0.06 | 0.9 | 0.6 | 18.5 |
| 2 | 0.5 | 7.5 | 1.5 | 9.0 |
| 3 | 5 | 75 | 6 | 3.1 |

**Figure 1**

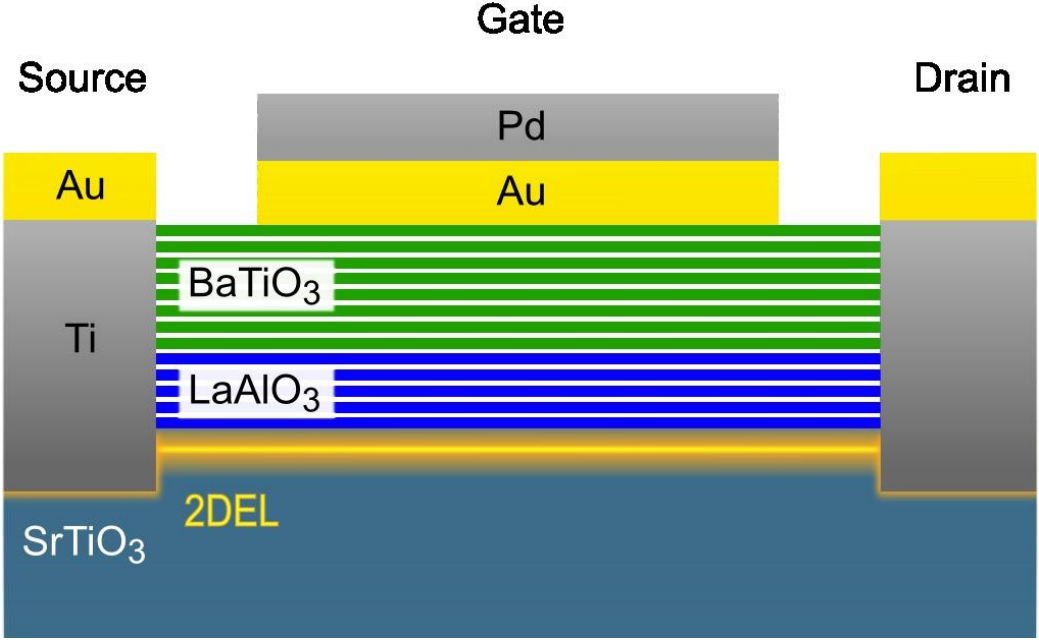

**Figure2**

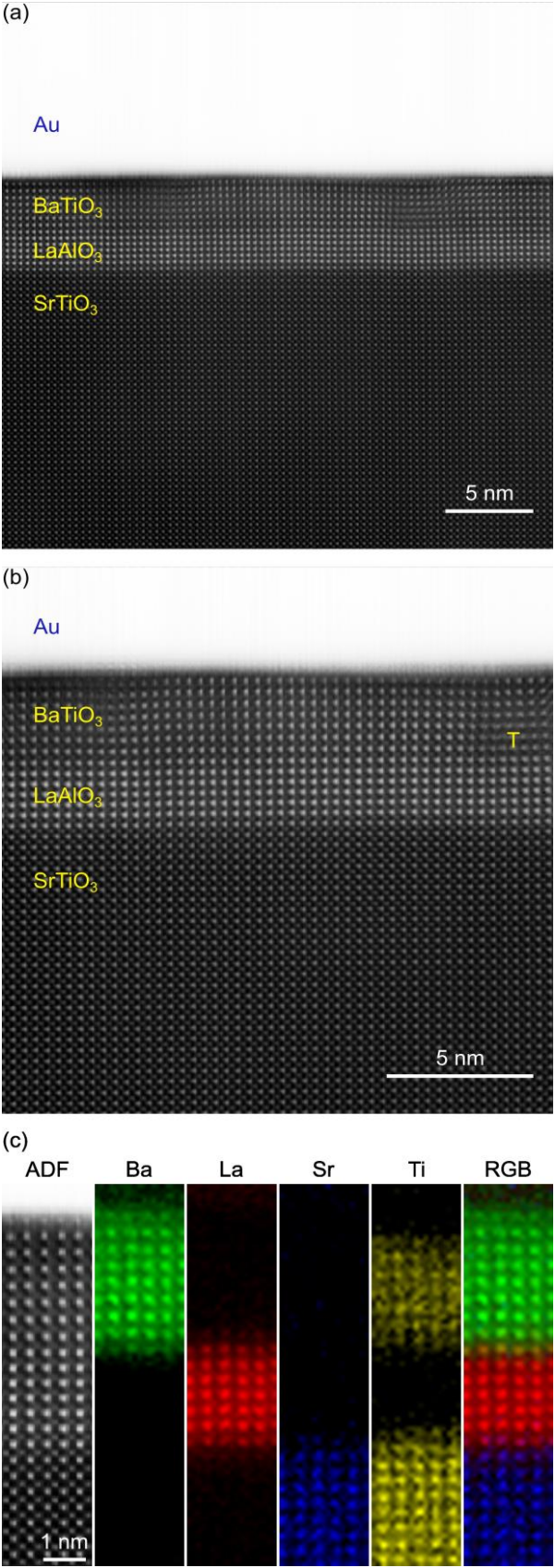

**Figure 3**

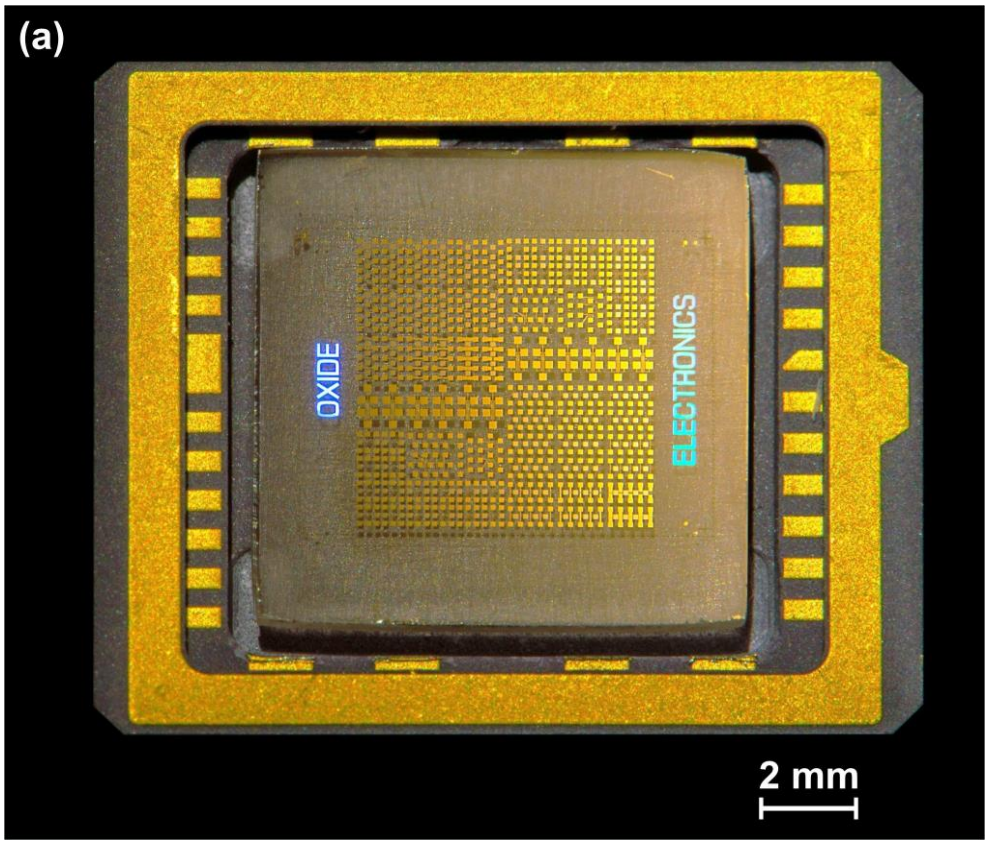

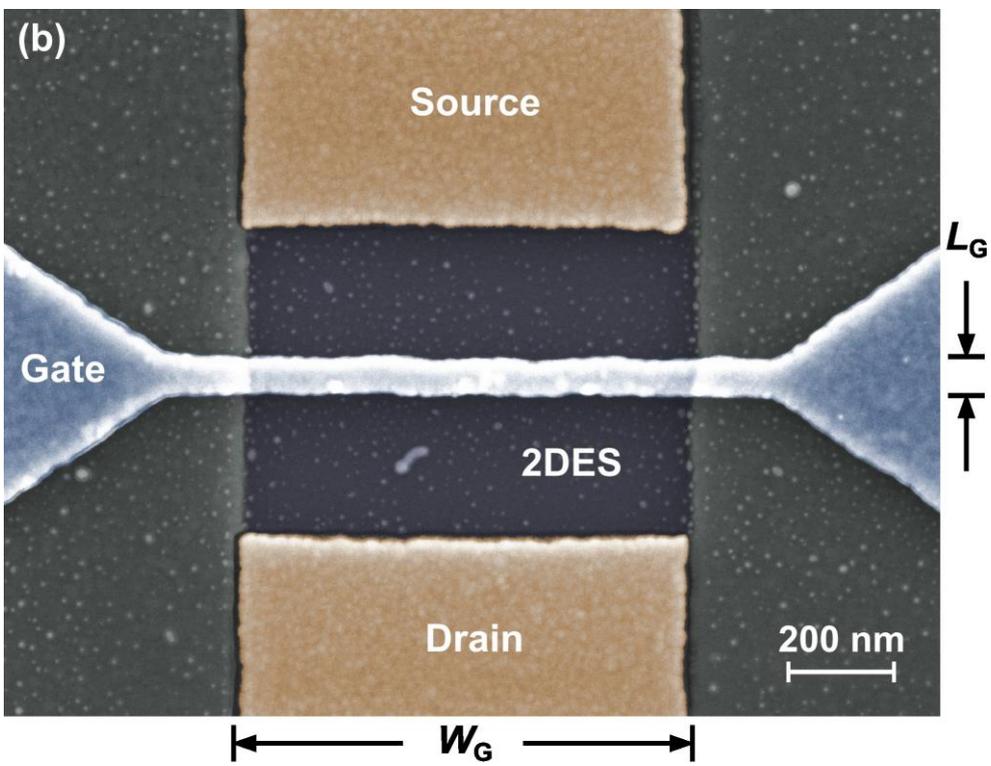

Figure 4

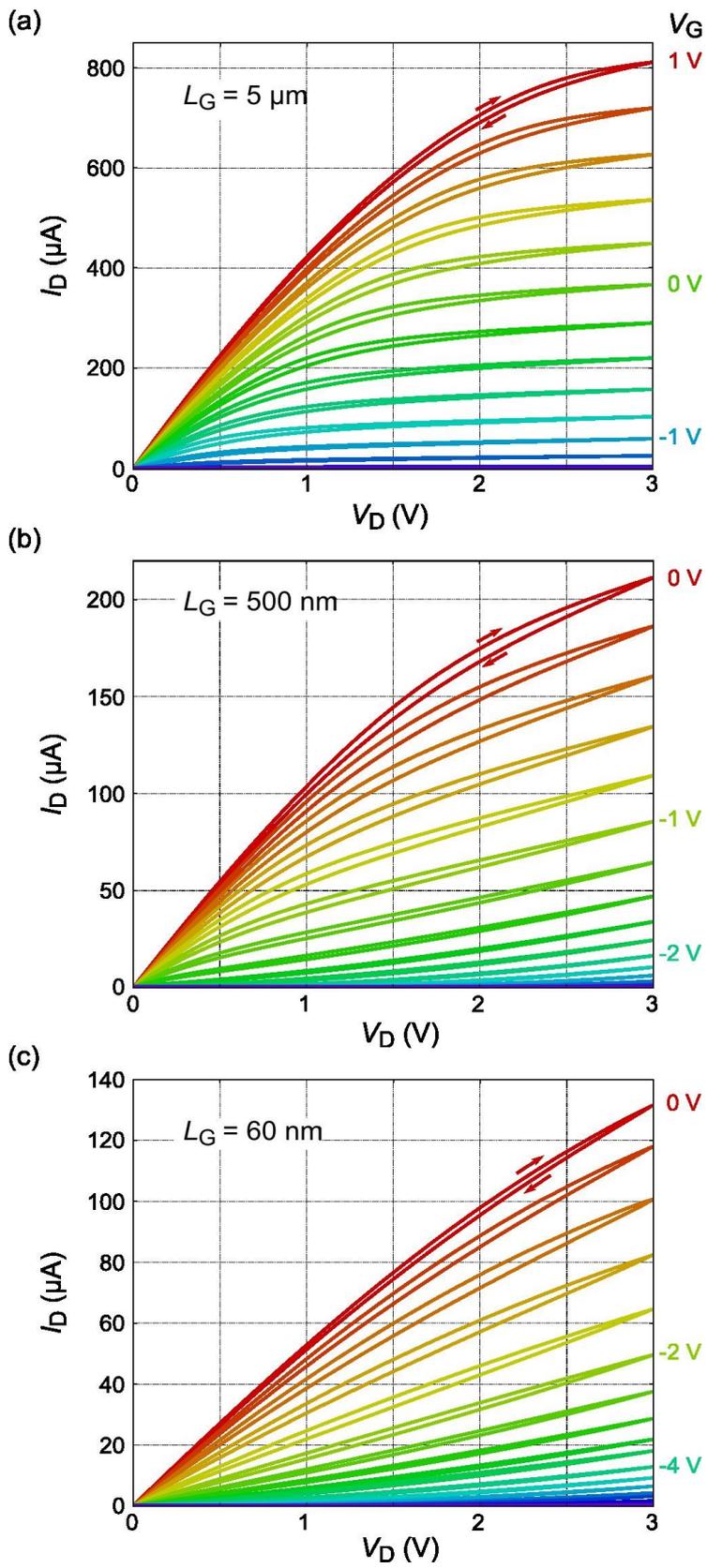

**Figure 5**

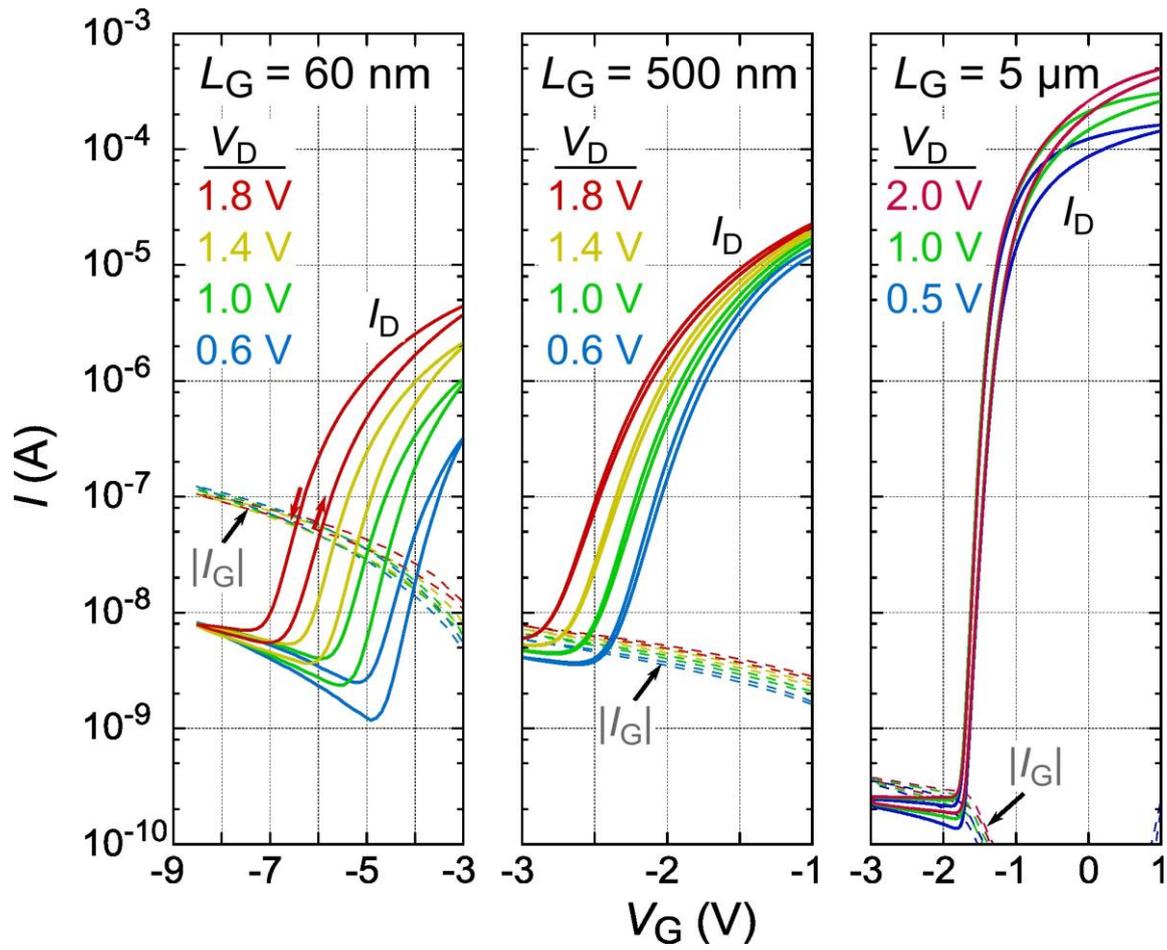